\documentclass[]{aa}

\usepackage{graphicx,multirow,booktabs,tabularx}
\usepackage{float}
\usepackage{txfonts}
\bibpunct{(}{)}{;}{a}{}{,}
\begin{document}

   \title{Long-term optical and radio variability of BL Lacertae}


   \author{Y. C. Guo \inst{1}
          \and
          S. M. Hu \inst{1,2}
          \and
          C. Xu \inst{1}
          \and
          C. Y. Liu \inst{1}
          \and
          X. Chen \inst{1,2}
          \and
          D. F. Guo \inst{1,2}
          \and
          F. Y. Meng \inst{1}
          \and
          M. T. Xu \inst{1}
          \and
          J. Q. Xu \inst{1}
          }

   \institute{School of Space Science and Physics, Shandong University, Weihai 264209, China\\
   \email{husm@sdu.edu.cn}
         \and
             Shandong Provincial Key Laboratory of Optical Astronomy and Solar-Terrestrial Environment, Institute of Space Sciences, Shandong University, Weihai 264209, China\\
             }

   \date{Received ; accepted }


  \abstract{Well-sampled optical and radio light curves of BL Lacertae in $B$, $V$, $R$, $I$ bands and 4.8, 8.0, 14.5 GHz from 1968 to 2014 were presented in this paper. A possible $1.26 \pm 0.05$ yr period in optical bands and a $7.50 \pm 0.15$ yr period in radio bands were detected based on discrete correlation function, structure function as well as Jurkevich method. Correlations among different bands were also analyzed and no reliable time delay was found between optical bands. Very weak correlations were detected between V band and radio bands. However, in radio bands the variation at low frequency lagged that at high frequency obviously. The spectrum of BL Lacertae turned mildly bluer when the object turned brighter, and stronger bluer-when-brighter trends were found for short flares. A scenario including a precessing helical jet and periodic shocks was put forward to interpret the variation characteristics of BL Lacertae.}


   \maketitle
%

\section{Introduction}

Blazars, namely, BL Lac objects and flat spectrum radio quasars, are a subclass of active galactic nuclei (AGNs). Blazars are characterized by several properties such as high luminosity, strong variability, polarization and non-thermal emission. Analyzing the variability of blazars is a good way to understand the nature of AGNs. The characteristic time scales of blazars range from minutes to years. Based on different time scales, the variability of blazars can be divided into three classes: intra-day variability (IDV), short-term variability (STV) and long-term variability (LTV). Variations over timescales ranging from a few minutes to less than one day are defined as IDV. Flux changes over a few days to less than one month are known as STV, while LTV refers to variations observed over months to several years \citep{G04,F05,G12}. Blazars commonly show random, significant outbursts, and the mechanism causing these outbursts are still in debate.

BL Lacertae is the prototype of BL Lac objects with a redshift z=0.069 \citep{M77}. Its host galaxy is an elliptical galaxy including a stellar population of about 0.7 Gyr in age \citep{H07}. BL Lacertae is a highly variable source whose extreme properties are thought to be attributed to its non-thermal emitting jet with relativistic velocity pointing close to the line of sight. It has been studied in detail by many intensive multi-frequency observations.

Quasi-periodic variations in optical and radio bands have been detected in many BL Lac objects. \citet{R01} recognized a ~5.7 yr period in the radio light curves of AO 0235+16. A $\sim$12 yr period was detected in the optical flux variation of OJ 287 \citep{V98,F10}. \citet{C14} analyzed the optical light curves of Mrk 421 and found a possible period of 1.36 yr. In particular, many authors have investigated the light curves of BL Lacertae for possible periods, leading to different results \citep{V04b}. A period of about 7.7 $\sim$ 7.8 yr in optical bands was verified by \citet{S95} and \citet{M96}. \citet{H02} found a 308 day period based on the optical light curves from 1980 to 1991. In radio bands, \citet{K03} reported a 1.4 yr period in 4.8 GHz, a period of 3.7 yr in 8.0 GHz and periods of 0.7, 1.6, 3.5 yr in 14.5 GHz flux variations. A period of about 8 yr was found by \citet{V04b} based on radio observations lasting 35 years.

Analysis on the correlations of flux variations among different frequencies could shed light on the inner emitting mechanism of BL Lacertae. \citet{V04a} interpreted the flux variability of BL Lacertae into two components: long-term mildly-chromatic variations and intraday flares characterized by a strong bluer-when-brighter chromatism. \citet{V09} found that the variation of average radio hardness ratio $H_{mean}$ lagged the optical outbursts by time scales varying from 100 to 300 days. Furthermore, a strong bluer-when-brighter behaviour was detected for inter-night variations, and the variations between different bands were correlated without precise time lag, except for the flare on 2005 September 6, which showed a lag of approximately 1052s between $B$ and $R$ band \citep{Z13}. \citet{R13} analyzed the correlation of BL Lacertae among flux variability in well-sampled optical and $\gamma-ray$ light curves of a flare started in May 2011, and a general correlation with likely no time lag was found.

In order to shed more light on the variability of BL Lacertae, long-term available optical and radio data were collected and analyzed. This paper is organized as follows: in Sect.~2, we present the long-term optical and radio light curves of BL Lacertae. The datasets in different bands are analyzed and possible periods are derived in Sect.~3, while correlations and spectral variability are discussed in Sect.~4 and Sect.~5, respectively. Results are presented in Sect.~6 and possible theoretical models which can explain variation characteristics of this object are discussed as well.

\section{Long-term data}

\begin{figure}
    \centering
     \includegraphics[width=\hsize]{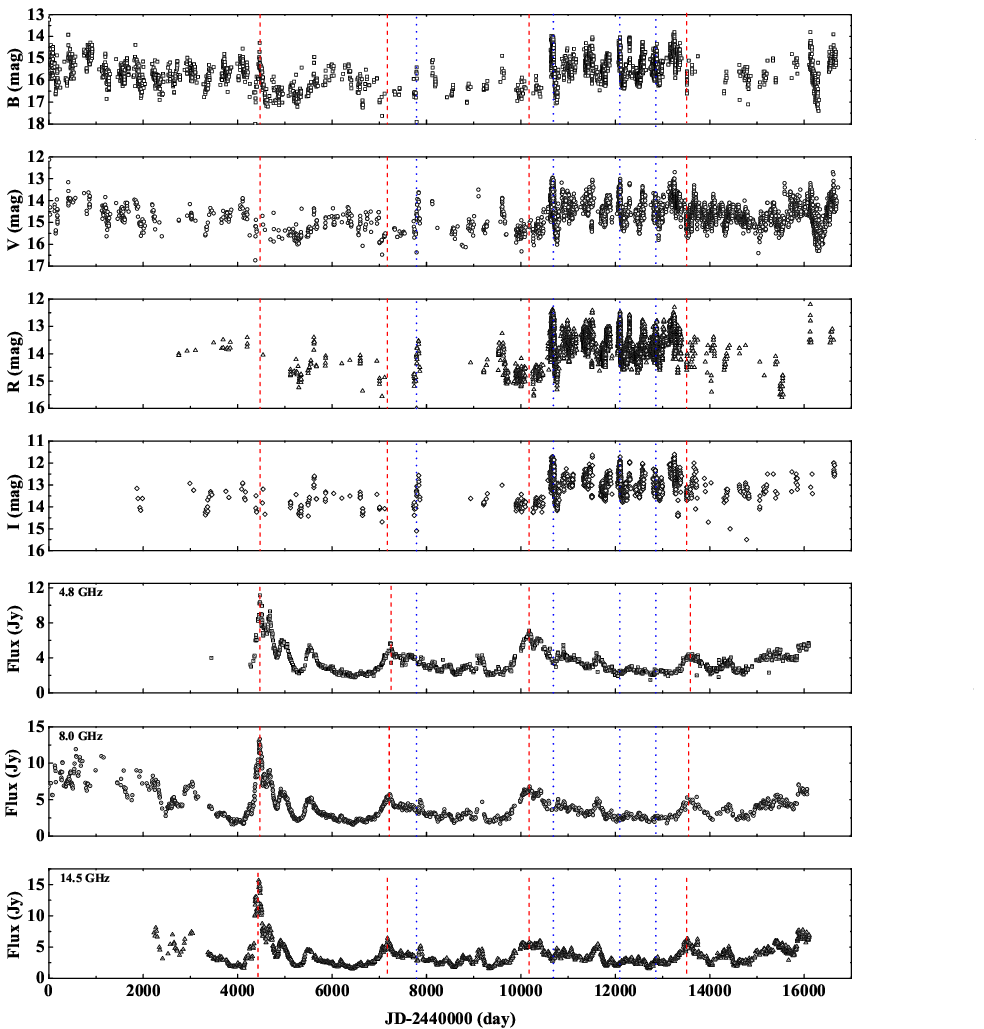}
      \caption{Optical and radio light curves of BL Lacertae. The top four panels are for $B$, $V$, $R$ and $I$ bands, while the bottom three panels are for 4.8, 8.0 and 14.5 GHz. The red dash lines show maxima of major radio flares. The blue dot lines label maxima of four major optical flares.}
         \label{or}
   \end{figure}

BL Lacertae is a typical blazar and has been observed by many groups. In this paper, we present the historical optical and radio data of BL Lacertae from literatures, Whole Earth Blazar Telescope (WEBT), Association of Variable Star Observers International Database (AAVSO), Steward Observatory spectropolarimetric monitoring project as well as University of Michigan Radio Astronomy Observatory (UMRAO).

Well-sampled optical light curves of BL Lacertae from April 1968 to February 2014 were presented in this paper. The majority of optical data were from WEBT, including results of observations performed by WEBT campaigns \citep{V02,V04b,V04a,V09,R09,R10} as well as data from literatures \citep{B99,C01,F00,F01,G00,G06,H04,W98,K00,M97,P03,S99,T99,Z04}. Some data points from January 1997 to February 2014 were obtained from AAVSO. Part of data points in $V$ band from October 2008 to July 2013 were obtained by Steward Observatory spectropolarimetric monitoring project \citep{S09}.

Radio data were acquired from UMRAO, performed with the 26m paraboloid at 4.8, 8.0 and 14.5 GHz. The observing details and data reduction were described by \citet{A85}.

The optical light curves in $B$, $V$, $R$ and $I$ bands are shown in the top four panels of Fig.~\ref{or}. The optical light curves appear as the superposition of fast variations on a long-term base level. The flux variations in all four bands are very strong and coincident with each other. Light curves in 4.8, 8.0 and 14.5 GHz radio bands are plotted in bottom three panels of Fig.~\ref{or}. Optical and radio behaviours appear quite different. The radio flux variations present as significant flares superposed on a long-term flux background with flux density close to zero. Red dash lines in Fig.~\ref{or} show maxima of four major radio flares at JD=2444474, 2447249, 2450173 and 2453510. These flares seem like a quasi-periodic event. The strongest flare happened at JD=2444474, which corresponded to a significant flare in $B$ band (maybe also in $V$, $R$ and $I$ bands). Whether the radio flare at JD=2447249 had an optical counterpart is unknown due to the lack of observation. The third radio flare happened at JD=2450173, when observations in $V$ band were also carried out, but no evidence for a optical counterpart was found. The radio flare at JD=2453510 had no optical counterparts in $B$, $V$, $R$ and $I$ bands. Blue dot lines in Fig.~\ref{or} illustrate maxima of four major optical flares happening at JD=2447215, 2449609, 2450772 and 2451401. The optical flares at JD=2447215, 2449609 and 2451401 were found corresponding to minor radio flares, and these radio flares became less and less significant from 14.5 GHz, 8.0 GHz to 4.8 GHz. The optical flare at JD=2450772 had no radio counterparts. Similar variation characteristics were detected for another BL Lac object S5 0716+714 by \citet{R03}, who found that the minor radio flux enhancements were simultaneous with the major optical outbursts.

\section{Periodicity analysis}

In this section we used the following methods to detect the periodicity of BL Lacertae in both optical and radio bands: discrete correlation function (DCF), first order structure function (SF) and Jurkevich method (JK).

In order to eliminate the influence of nonuniformity, all optical data were firstly binned by one day, then the data in $B$ and $V$ bands were binned by 5 days. We primarily applied these statistical tools to light curves in $B$ and $V$ bands, and the results were then checked by light curves in $R$ and $I$ bands, because the light curves in $B$ and $V$ bands were comparatively better distributed.

\begin{figure}
   \centering
   \includegraphics[width=\hsize]{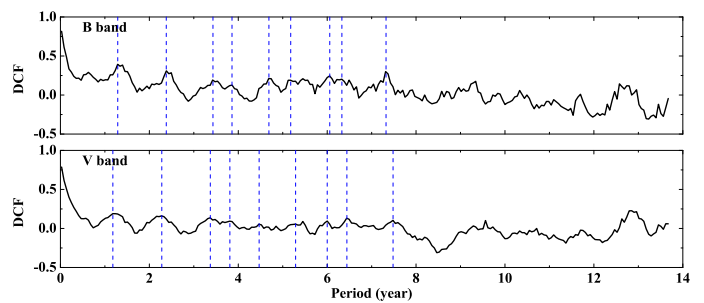}
      \caption{Period analysis results by DCF in $B$ (the top panel) and $V$ (the bottom panel) bands for BL Lacertae. The dash lines present the possible periods listed in Col.~2 and Col.~3 of Tab.~\ref{optical_tab}.}
         \label{dcf_optical}
   \end{figure}

The discrete correlation function is intended for unevenly-sampled astronomical datasets, and it was described in detail by \citet{E88}. It turned out to be excellent for single-period analysis, but not for multiple periods \citep{F06}. DCF was applied to the optical light curves and the results in $B$ and $V$ bands were shown in Fig.~\ref{dcf_optical}. Peaks of DCF imply possible periods. The possible periods in $B$ and $V$ bands are listed in Col.~2 and Col.~3 of Tab.~\ref{optical_tab} and marked by blue dash lines in Fig.~\ref{dcf_optical}. It is noticeable that $P_2$, $P_4$, $P_6$, $P_8$ and $P_9$ are likely to be multiples of $P_1$.

\begin{figure}
   \centering
   \includegraphics[width=\hsize]{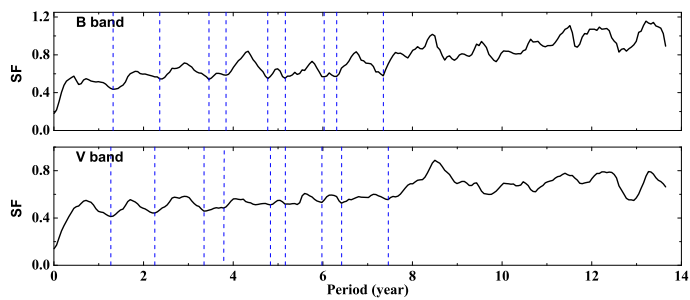}
      \caption{Results by SF in $B$ (the top panel) and $V$ (the bottom panel) bands for BL Lacertae. The possible periods listed in Col.~4 and Col.~5 of Tab.~\ref{optical_tab} are marked by dash lines.}
         \label{sf_optical}
   \end{figure}

Then we used SF, a powerful and well-studied method for period analysis \citep{S85,P97,G08}. Application of SF led to a periodogram shown in Fig.~\ref{sf_optical}. The minima of SF indicate possible periods, while the maxima demonstrate possible time scales. It can be seen that the results for $B$ and $V$ bands are in consistent with each other. Possible periods are presented in Col.~4 and Col.~5 of Tab.~\ref{optical_tab}. $P_2$, $P_4$, $P_6$, $P_8$ and $P_9$ are likely to be two, three, four, five and six times of $P_1$, respectively.

\begin{figure}
   \centering
   \includegraphics[width=\hsize]{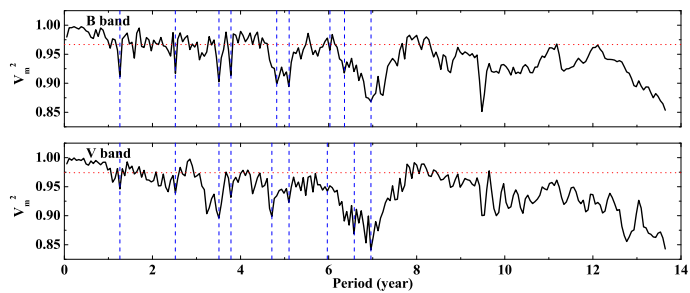}
      \caption{Results by Jurkevich method in $B$ and $V$ bands. The top and bottom panels show results derived from $B$ and $V$ bands, respectively. The horizontal dot lines represent FAP level of 0.01. The vertical dash lines label the possible periods listed in Col.~6 and Col.~7 of Tab.~\ref{optical_tab}.}
         \label{jurk_optical}
   \end{figure}

The Jurkevich method \citep{J71} is based on the expected mean square of deviation, and can probe periods without producing fake periods \citep{F10}. Normalized $V{_m}^2$ for light curves in $B$ and $V$ bands are shown in Fig.~\ref{jurk_optical}, and the minima of $V{_m}^2$ indicate possible periods. To confirm the existence of periods, we adopted the False Alarm Probability (FAP) to provide a quantitative criterion \citep{H86,F10}. The horizontal dash lines in Fig. \ref{jurk_optical} represent FAP level of 0.01, namely, significance level of 0.99. The possible periods are listed in Col.~6 and Col.7 of Tab.~\ref{optical_tab}. It can be seen that $P_2$, $P_4$, $P_6$, $P_8$ and $P_9$ are possible multiples of $P_1$.

DCF, SF and JK were applied to the optical light curves of BL Lacertae. Different bin sizes were used in the calculation and the results showed no significant difference. All results are shown in Tab.~\ref{optical_tab} and confirmed a possible period of $1.26 \pm 0.05$ yr, because this period as well as its harmonics $P_2$, $P_4$, $P_6$, $P_8$ and $P_9$ are apparent. Possible periods of $3.44 \pm 0.07$, $4.72 \pm 0.13$ and $6.01 \pm 0.03$ yr were also probed. All possible periods are also obvious for light curves in $R$ and $I$ bands by the application of above-mentioned methods.

\begin{table*}
\caption{Results of period analysis for $B$ and $V$ bands. All periods and their standard deviation (in brackets) are in unit of year.}
\label{optical_tab}
\centering
\begin{tabular}{*{8}{c}}
\hline\hline
\multirow{2}*{Period} & \multicolumn{2}{c}{DCF} & \multicolumn{2}{c}{SF} & \multicolumn{2}{c}{JK} & \multirow{2}*{Mean}\\
\cmidrule(lr){2-3}\cmidrule(lr){4-5}\cmidrule(lr){6-7}
& B & V & B & V & B & V & \\
\midrule
1 & 1.29 & 1.18 & 1.32 & 1.27 & 1.26 & 1.26 & 1.26 (0.05) \\
2 & 2.38 & 2.28 & 2.36 & 2.25 & 2.52 & 2.52 & 2.39 (0.12) \\
3 & 3.43 & 3.37 & 3.46 & 3.35 & 3.51 & 3.51 & 3.44 (0.07) \\
4 & 3.86 & 3.81 & 3.84 & 3.79 & 3.78 & 3.78 & 3.81 (0.03) \\
5 & 4.69 & 4.47 & 4.77 & 4.83 & 4.82 & 4.71 & 4.72 (0.13) \\
6 & 5.18 & 5.29 & 5.16 & 5.16 & 5.10 & 5.10 & 5.17 (0.07) \\
7 & 6.06 & 6.00 & 6.03 & 5.98 & 6.03 & 5.97 & 6.01 (0.03) \\
8 & 6.33 & 6.44 & 6.31 & 6.42 & 6.36 & 6.58 & 6.41 (0.10) \\
9 & 7.32 & 7.48 & 7.35 & 7.46 & 6.96 & 6.96 & 7.26 (0.24) \\
\hline
\end{tabular}
\end{table*}

\begin{figure}
   \centering
   \includegraphics[width=\hsize]{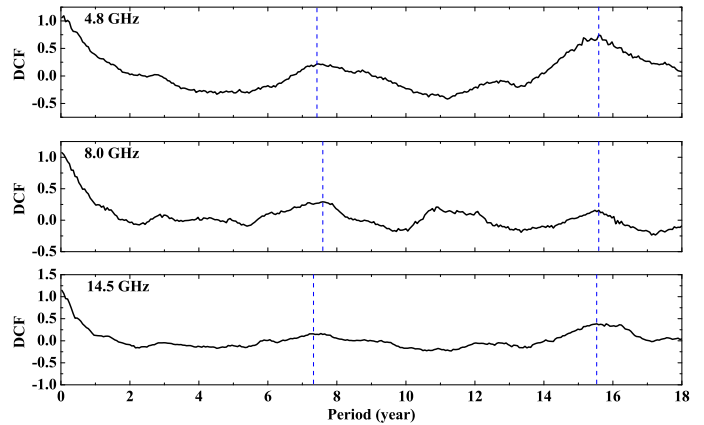}
      \caption{DCF results for light curves in 4.8 (the top panel), 8.0 (the middle panel) and 14.5 (the bottom panel) GHz for BL Lacertae. Blue dash lines refer to possible periods listed in Col.~2, Col.~3 and Col.~4 of Tab.~\ref{radio_tab}}
         \label{dcf_radio}
   \end{figure}

\begin{figure}
   \centering
   \includegraphics[width=\hsize]{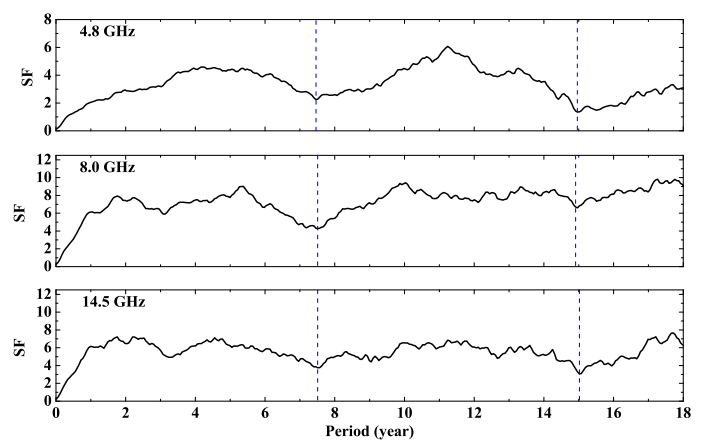}
      \caption{SF results for BL Lacertae in radio bands. The upper, middle, and lower panels demonstrate results derived from data in 4.8, 8.0 and 14.5 GHz, respectively. Possible periods in Col.~5, Col.~6 and Col.~7 of Tab.~\ref{radio_tab} are marked by blue dash lines.}
         \label{sf_radio}
   \end{figure}

\begin{figure}
   \centering
   \includegraphics[width=\hsize]{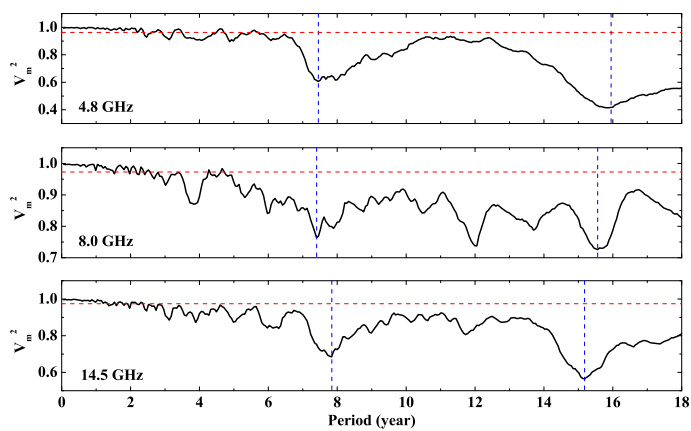}
      \caption{Results of Jurkevich method in radio bands. Three panels from top to bottom are for 4.8, 8.0 and 14.5 GHz, respectively. Horizontal dash lines represent FAP level of 0.01. Vertical dash lines mark possible periods in Col.~8 Col.~9 and Col.~10 of Tab.~\ref{radio_tab}.}
         \label{jurk_radio}
   \end{figure}

\begin{table*}
\caption{Results of period analysis for radio bands. All periods and their standard deviation (in brackets) are in unit of year.}
\label{radio_tab}
\centering
\begin{tabular}{*{11}{c}}
\hline\hline
\multirow{2}*{Period} &\multicolumn{3}{c}{DCF} & \multicolumn{3}{c}{SF} & \multicolumn{3}{c}{JK} & \multirow{2}*{Mean}\\
\cmidrule(lr){2-4}\cmidrule(lr){5-7}\cmidrule(lr){8-10}
& 4.8 GHz & 8.0 GHz & 14.5 GHz & 4.8 GHz & 8.0 GHz & 14.5 GHz & 4.8 GHz & 8.0 GHz & 14.5 GHz & \\
\midrule
1 & 7.42  & 7.59  & 7.32  & 7.46  & 7.51  & 7.51  & 7.45  & 7.4   & 7.84  & 7.50 (0.15) \\
2 & 15.59 & 15.59 & 15.53 & 14.96 & 14.91 & 15.02 & 15.95 & 15.56 & 15.18 & 15.37 (0.36) \\
\hline
\end{tabular}
\end{table*}

The same analysis methods were also performed on radio light curves obtained from UMRAO. Fig.~\ref{dcf_radio}, Fig.~\ref{sf_radio} and Fig.\ref{jurk_radio} present the results derived by DCF, SF and JK, respectively. The two strongest signals are $7.50 \pm 0.15$ yr and its harmonic $15.37 \pm 0.36$ yr. All results are listed in Tab.~\ref{radio_tab}, as well as labeled by blue dash lines in Fig.~\ref{dcf_radio}, Fig.~\ref{sf_radio} and Fig.\ref{jurk_radio}. This $7.50 \pm 0.15$ yr period is also in accordance with the time intervals between maxima of flares in radio bands labeled by red dash lines in Fig.~\ref{or}. The $1.26 \pm 0.05$ yr period detected in optical bands was not probed in radio bands.

\section{Correlation analysis}

The analysis of a time delay between different bands is very important, because it can provide useful information for the emission mechanism. The correlations between optical and radio bands are of particular importance, which are both attributed to synchrotron radiation \citep{R03}. In this section, DCF was used to detect the correlation between different bands \citep{E88}. Two sets of data presenting the flux variations of two different bands were imported into discrete correlation function, and positive lags of DCF indicated that the variation of the second band delayed the first band by the lag calculated.

\subsection{Optical-optical correlation}

\begin{figure}
   \centering
   \includegraphics[width=\hsize]{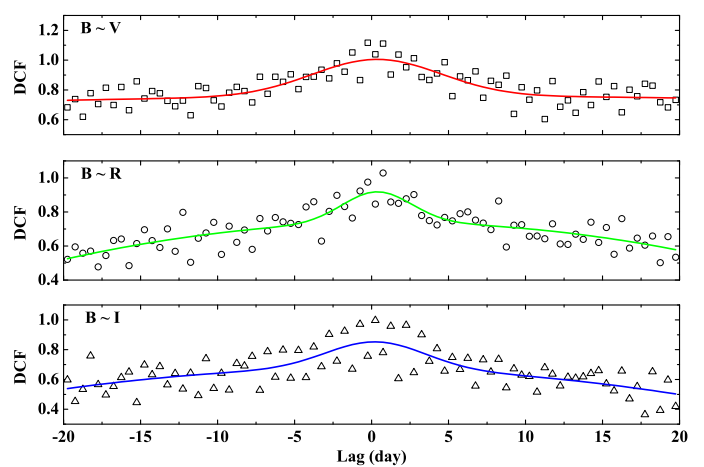}
      \caption{Results by DCF in different optical bands. Open squares in the top panel illustrate the results of DCF between $B$ and $V$ band. Circles in the middle panel demonstrate correlation between $B$ and $R$ band, while triangles in the bottom panel are for $B$ and $I$ band. Solid lines present the results of Gaussian Fitting.}
         \label{oo}
   \end{figure}

According to the light curves shown in Fig. \ref{or}, it is evident that the variations of all optical bands follow the same trend, and flares emerge roughly simultaneously. The results of DCF analysis are presented in Fig.~\ref{oo}. Each peak of DCF is broad, and hardly can a precise lag be determined, so Gaussian Fitting was applied to determine the centre. The results showed that flux variations in $B$ band may be delayed with respect to those in $V$, $R$ and $I$ bands by $0.330 \pm 0.468$, $0.343 \pm 0.421$ and $0.262 \pm 0.723$ day, respectively. The computation of Gaussian Fitting was of big uncertainty, and the derived delay was less than the resolution of DCF (0.5 day). Furthermore, it was meaningless to improve the resolution of lag due to the lack of better-sampled light curves. Therefore, good correlation was found between optical bands, but we could not derive a precise and reliable lag smaller than 0.5 day. \citet{V02} and \citet{W14} also investigated the correlation between light curves of BL Lacertae in different optical bands, and no significant and measurable time delay was found. Our results indicate that there are no reliable time delays between different optical bands.

\subsection{Optical-radio correlation}

\begin{figure}
   \centering
   \includegraphics[width=\hsize]{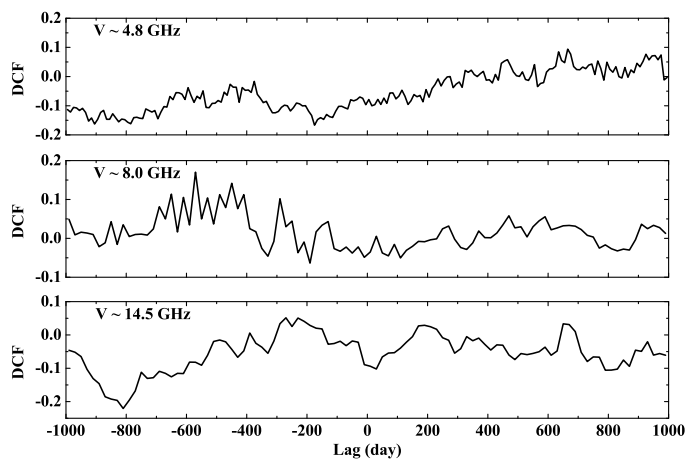}
      \caption{Results of DCF between $V$ and radio bands. The top panel represents the result of DCF between $V$ band and 4.8 GHz. The middle panel demonstrates the correlation between $V$ band and 8.0 GHz, while the bottom panel is for $V$ band and 14.5 GHz.}
         \label{vr}
   \end{figure}

Possible time delay between $V$ band and radio bands was also investigated. The magnitude in $V$ band was converted to flux density using the zero magnitude equivalent flux density given by \citet{M90}. Fig. \ref{vr} shows the results of DCF. Very weak correlation was found between optical and radio bands. There is no positive peak at zero time lag, which illustrates the flux variation in optical band has no obvious and simultaneous duplicates in radio bands. The weakness of DCF peaks leads to the conclusion that there is no reliable correlation between $V$ and radio bands for BL Lacertae. Similar result was also presented by \citet{R03} for BL Lac object S5 0716+714.

\subsection{Radio-radio correlation}

\begin{figure}
   \centering
   \includegraphics[width=\hsize]{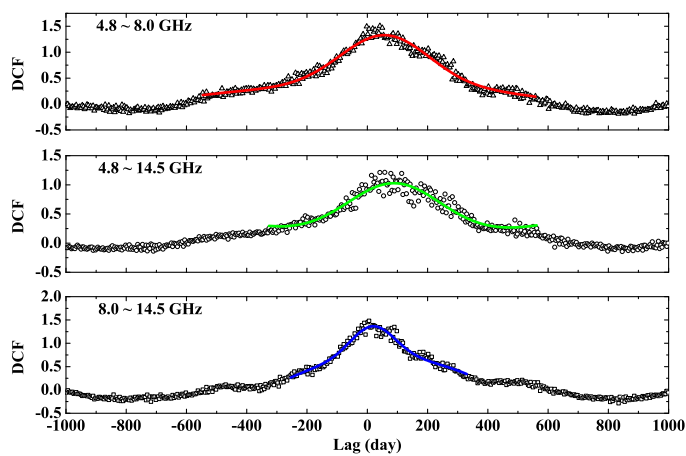}
      \caption{Time lags of flux variations between different radio bands by DCF. Squares in the top panel present DCF between 4.8 and 8.0 GHz, circles in the middle panel display time lag between 4.8 and 14.5 GHz, and triangles in the bottom panel illustrate the correlation between 8.0 and 14.5 GHz. Solid lines represent the results of Gaussian Fitting.}
         \label{rr}
   \end{figure}

The radio bands reveal similar behaviour according to the light curves shown in Fig. \ref{or}. DCF was also used to detect the correlation between radio bands, and the results were shown in Fig. \ref{rr}. The variation at 4.8 GHz lags that at 8.0 GHz by $58.2 \pm 2.3$ days and lags that at 14.5 GHz by $87.2 \pm 3.2$ days, while the variation at 14.5 GHz leads that at 8.0 GHz by $23.8 \pm 2.1$ days. It is noticeable that the delay bewteen 4.8 and 8.0 GHz plus the delay between 8.0 and 14.5 GHz approximately equals to the delay between 4.8 and 14.5 GHz within error, and the variation at higher frequency leads that at lower frequency.

\section{Spectral variability}

\begin{figure}
    \centering
    \includegraphics[width=\hsize]{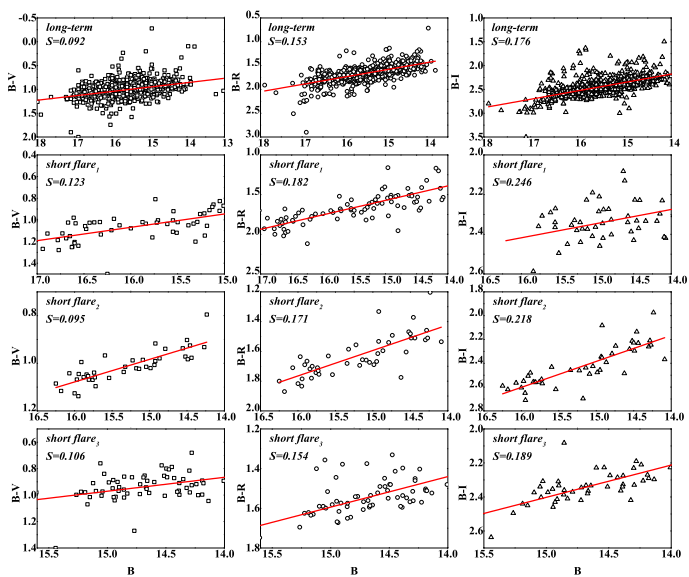}
      \caption{Correlation between color index and brightness. Squares in the left four panels illustrate the relationship between $B$ and $B-V$, circles in the middle panels present the relationship between $B$ and $B-R$, and triangles in the right four panels display the relationship between $B$ and $B-I$. The top three panels demonstrate the long-term trends, while the others reflect the trends of three short flares. The solid lines represent the best linear fits, and the color variability indicator $S$ is labeled in each panel.}
         \label{color}
   \end{figure}

\begin{table*}
\caption{The results of color analysis.}
\label{colortab}
\centering
\begin{tabular}{*{6}{c}}     
\hline\hline
$color~index$        & $type$    & $JD$              & $S$   & ${\Delta}S$ & $mean$                   \\
\midrule
\multirow{4}{*}{$B-V$} & $long-term$ & 2439974 $\sim$ 2456676 & 0.092 &                   &                        \\ \cmidrule(lr){5-6}
                     & $flare_1$    & 2450627 $\sim$ 2450782 & 0.123 & 0.031             & \multirow{3}{*}{0.016} \\
                     & $flare_2$    & 2452088 $\sim$ 2452138  & 0.094 & 0.002             &                        \\
                     & $flare_3$    & 2453191 $\sim$ 2453284 & 0.106 & 0.013             &                        \\ \midrule
\multirow{4}{*}{$B-R$} & $long-term$ & 2442753 $\sim$ 2456624 & 0.153 &                   &                        \\ \cmidrule(lr){5-6}
                     & $flare_1$    & 2450627 $\sim$ 2450782 & 0.182 & 0.030             & \multirow{3}{*}{0.017} \\
                     & $flare_2$    & 2452088 $\sim$ 2452138  & 0.171 & 0.019             &                        \\
                     & $flare_3$    & 2453191 $\sim$ 2453284 & 0.154 & 0.001             &                        \\ \midrule
\multirow{4}{*}{$B-I$} & $long-term$ & 2441867 $\sim$ 2456643 & 0.176 &                   &                        \\ \cmidrule(lr){5-6}
                     & $flare_1$    & 2450627 $\sim$ 2450781 & 0.246 & 0.070             & \multirow{3}{*}{0.042} \\
                     & $flare_2$    & 2452088 $\sim$ 2452138  & 0.218 & 0.042             &                        \\
                     & $flare_3$    & 2453191 $\sim$ 2453284 & 0.189 & 0.014             &                        \\
\hline
\end{tabular}
\end{table*}

The spectral variability of BL Lacertae with respect to its brightness was analyzed in this section. Many authors have done researches on this topic. The results of \citet{R70} and \citet{V03} all came to a conclusion that BL Lacertae tended to be bluer when brighter. Optical data of $B$, $V$, $R$ and $I$ bands were binned by one day, then data points in the same day were taken as quasi-simultaneous data. Color indexes, $B-V$, $B-R$ and $B-I$ were calculated. In order to compare the color variability on long-term trend and short flares with a time scale of tens of days, we chose three short flares with enough quasi-simultaneous data points. The durations of the three short flares are: $short flare_1$ $(2450627 \sim 2450782)$, $short flare_2$ (2452088 $\sim$ 2452138) and $short flare_3$ (2453191 $\sim$ 2453284). The linear fitting slope $S$ of color index versus $B$ magnitude was taken as the color variability indicator \citep{H11}. The color variability indicators of long-term variability and short flares were derived and compared.

The results are shown in Fig.~\ref{color}. Panels in the left column illustrate the relationship between B and $B-V$, while panels in the middle and right columns show the trends of $B$ varying with $B-R$ and $B-I$, respectively. Panels in the four rows from top to bottom display the trends of long-term variation, $short flare_1$, $short flare_2$ and $short flare_3$, respectively. The label marking long-term trend or short flares, as well as the color variability indicator $S$ are given in each panel. Positive correlations between color index and brightness were found for all cases. All results are listed in Tab.~\ref{colortab}. In Tab. \ref{colortab}, Col.~1 presents the names of color indexes; Col.~2 labels whether the calculation is based on long-term data or short flares; Col.~3 is the time span of long-term data set or short flares; Col.~4 presents color variability indicator $S$; Col.~5 lists ${\Delta}S=S_{flare}-S_{long-term}$; Col.~6 refers to the average of ${\Delta}S$ for each color index. According to the results, $S$ of short flares are commonly larger than that of long-term variation, and it is interesting that ${\Delta}S$ becomes larger from $B-V$, $B-R$ to $B-I$. So we conclude that the spectrum turns mildly bluer when the source turns brighter, and the spectral variability indicator is bigger on short flares. This difference between long-term trend and short flares becomes more evident with wider spectral span. Our result is in accordance with the result of \citet{V04a}, who interpreted the variability of BL Lacertae in terms of a ``mildly-chromatic'' longer-term trend as well as a ``strongly-chromatic'' shorter-term one.

\section{Summary and discussion}

Well-sampled historical optical light curves in $B$, $V$, $R$ and $I$ bands and radio light curves in 4.8, 8.0 and 14.5 GHz of BL Lacertae were analyzed, and our main results are as follows:

\begin{itemize}
\item Optical and radio behaviours of BL Lacertae appeared different. The flares in optical bands commonly corresponded to minor or even no radio flares, which became more inconspicuous with frequency decreasing. Some major flares in radio bands had no optical counterparts, but the greatest radio flare which reached its maximum at JD=2444474 corresponded to a significant optical flare.
\item Possible periods of $1.26 \pm 0.05$, $3.44 \pm 0.07$, $4.72 \pm 0.13$ and $6.01 \pm 0.03$ yr were detected in optical bands by three methods. Among all these possible periods detected in optical bands, the $1.26 \pm 0.05$ yr period is the surest one. And a clearer evidence of a $7.50 \pm 0.15$ yr period was found from the radio light curves in 4.8, 8.0 and 14.5 GHz, in accordance with the $\sim 8$ year period found by \citet{V04b}.
\item Good correlations were found for variations in optical bands at different frequencies. No reliable time delay could be determined by DCF. The variations in radio bands were well-correlated too, and the application of DCF confirmed the existence of time delay: the variation at 4.8 GHz lagged those at 8.0 GHz and 14.5 GHz by $58.2 \pm 2.3$ days and $87.2 \pm 3.2$ days, respectively, and the variation at 14.5 GHz led that at 8.0 GHz by $23.8 \pm 2.1$ days. However, very weak correlation was found between V band and radio bands.
\item For long-term variation, the spectrum of BL Lacertae turned mildly bluer when the object turned brighter, and stronger bluer-when-brighter trends were found in short flares. With a wider span of spectrum, the bluer-when-brighter trend in flares became more evident.
\end{itemize}

The unified model of AGNs includes a supermassive black hole in the center of host galaxy and an accretion disk surrounding it. Periods in flux variability of blazars may be caused by ballistic or nonballistic helical trajectories in jets, which leads to a periodically changing viewing angle. According to \citet{R04}, the observed period of nonballistic helical motion is shorter than the real physical driving period due to light-travel time effects. \citet{R04} also discussed the possible range of period for three nonballistic periodic driving mechanisms: (1) observed periods $P_{obs} \lesssim 10$ days for massive quasars, or $P_{obs}$ on the order of 1 day for BL Lac objects, may be attributed to internal jet rotation, (2) periodicity due to orbital-driven helical motion in a binary black hole system (BBHS) is usually constrained to $P_{obs} \gtrsim 10$ days, and (3) Newtonian-driven precession in a ballistic or nonballistic helical jet caused by a close BBHS can account for $P_{obs} \gtrsim 1$ year. The periods we detected are all in the order of magnitudes of 1 year, but the difference between periods detected in optical and radio bands is confusing and inspiring.

Different behaviors in optical and radio variations were unfolded, and the correlation between optical and radio flux variations was very weak. \citet{V04b} introduced a scenario that the mechanisms leading to optical outbursts could also affect the lower energy flux variations. Therefore, the correlation and the possible delay between optical and radio variation became less evident with the radio frequency decreasing. \citet{R03} also derived a similar result from the long-term behaviour of another BL Lac object S5 0716+714, who believed that the mechanism responsible for optical outbursts could only marginally affect the radio bands, and another process not affecting the optical bands led to the dominant radio events.

Assuming that the jet precesses at a certain angular velocity, we can hardly use a single ballistic or nonballistic helical jet model to account for different periods detected in the optical and radio bands. Therefore, we consider that the flux variation of BL Lacertae can be roughly interpreted by a precessing helical jet with periodic shocks. In this scenario, periodic shocks trigger the optical variation in a period of $1.26 \pm 0.05$ yr produced at inner denser parts of the jet, but the affect of shocks is insignificant when they are propagated to the more external region where radio emission generated. Therefore, the $7.50 \pm 0.15$ yr period found in the radio flux variations may be caused by the precessing jet. Because of the lighthouse effect of precessing jet, the radio flux variation appears as flares superposed on a long-term background with its flux density close to zero . We also found that the radio variations at higher frequencies led those at lower frequencies obviously, and the lag was larger with the frequency interval elongating. The viewing angle changes with helical jet movement leads to the change of Doppler factor, which results in the lags between different radio bands inherently, because different frequency emitting portions of the jet need the same viewing angle \citep{V99,R03}. On the other hand, this scenario could typically describe the strong bluer-when-brighter phenomena we found in the optical flares, because the electron cooling mechanism leads to larger amplitude at higher frequency without any delay \citep{W07,W14}. \citet{V09} detected the correlations between the optical flux densities and the average radio hardness ratio $H_{mean}$, and found that the hard radio events lagged behind the optical outbursts with time scales varying from 100 to 300 days. This phenomena could also be interpreted by our scenario: although the shocks become negligible in the radio flux emitting region, they still make the spectrum harder.

In this paper, long-term and well-sampled optical and radio light curves of BL Lacertae from April 1968 to February 2014 were presented. Then periodicity analysis methods were applied to the light curves to detect possible periods. Correlations between different frequencies were detected and long-term optical spectral variability was analyzed as well. Finally a proper scenario was put forward to describe variable characteristics of this object. This scenario could explain the long-term optical and radio behaviour of BL Lacertae fairly well, but it is still worth inquiring that very weak correlation was found between optical and radio flux variation. More well-sampled long-term observations and studies are needed for further understanding of the radiation mechanism.

\begin{acknowledgements}
This work is supported by the National Natural Science Foundation of China under grants No. 11203016, 11143012, and by the National Natural Science Foundation of China and Chinese Academic of Sciences joint fund on astronomy under project No. 10778701, 10778619. This research has made use of data from the University of Michigan Radio Astronomy Observatory which has been supported by the University of Michigan and by a series of grants from the National Science Foundation, most recently AST-0607523. We also thanks for data taken and assembled by the WEBT collaboration and stored in the WEBT archive at the Osservatorio Astrofisico di Torino - INAF (\url{http://www.oato.inaf.it/blazars/webt/}), data from the Steward Observatory spectropolarimetric monitoring project supported by Fermi Guest Investigator grants NNX08AW56G, NNX09AU10G, and NNX12AO93G as well as data from the variable star observations from the AAVSO International Database contributed by observers worldwide.
\end{acknowledgements}
\bibliographystyle{aa} 
\bibliography{reference} 

\begin{thebibliography}{53}
\expandafter\ifx\csname natexlab\endcsname\relax\def\natexlab#1{#1}\fi

\bibitem[{{Aller} {et~al.}(1985){Aller}, {Aller}, {Latimer}, {et~al.}}]{A85}
{Aller}, H.~D., {Aller}, M.~F., {Latimer}, G.~E., {et~al.} 1985, \apjs, 59, 513

\bibitem[{{Bai} {et~al.}(1999){Bai}, {Xie}, {Li}, {et~al.}}]{B99}
{Bai}, J.~M., {Xie}, G.~Z., {Li}, K.~H., {et~al.} 1999, \aaps, 136, 455

\bibitem[{{Chen} {et~al.}(2014){Chen}, {Hu}, {Guo}, \& {Du}}]{C14}
{Chen}, X., {Hu}, S.~M., {Guo}, D.~F., \& {Du}, J.~J. 2014, Ap{\&}SS, 349, 909

\bibitem[{{Clements} \& {Carini}(2001)}]{C01}
{Clements}, S.~D. \& {Carini}, M.~T. 2001, \aj, 121, 90

\bibitem[{{Edelson} \& {Krolik}(1988)}]{E88}
{Edelson}, R.~A. \& {Krolik}, J.~H. 1988, \apj, 333, 646

\bibitem[{{Fan}(2005)}]{F05}
{Fan}, J.~H. 2005, ChJAS, 5, 213

\bibitem[{{Fan} \& {Lin}(2000)}]{F00}
{Fan}, J.~H. \& {Lin}, R.~G. 2000, \apj, 537, 101

\bibitem[{{Fan} {et~al.}(2010){Fan}, {Liu}, {Qian}, {Tao}, {Shen}, {Zhang},
  {Huang}, \& {Wang}}]{F10}
{Fan}, J.~H., {Liu}, Y., {Qian}, B.~C., {et~al.} 2010, RAA, 10, 1100

\bibitem[{{Fan} {et~al.}(2001){Fan}, {Qian}, \& {Tao}}]{F01}
{Fan}, J.~H., {Qian}, B.~C., \& {Tao}, J. 2001, \aap, 369, 758

\bibitem[{{Fan} {et~al.}(2006){Fan}, {Tao}, {Qian}, {Gupta}, {Liu}, {Yuan},
  {Yang}, {Wang}, \& {Huang}}]{F06}
{Fan}, J.~H., {Tao}, J., {Qian}, B.~C., {et~al.} 2006, \pasj, 58, 797

\bibitem[{{Gaur} {et~al.}(2012){Gaur}, {Gupta}, {Strigachev}, {Bachev},
  {Semkov}, {Wiita}, {Peneva}, {Boeva}, {Slavcheva-Mihova}, {Mihov}, {Latev},
  \& {Pandey}}]{G12}
{Gaur}, H., {Gupta}, A.~C., {Strigachev}, A., {et~al.} 2012, \mnras, 425, 3002

\bibitem[{{Ghosh} {et~al.}(2000){Ghosh}, {Ramsey}, {Sadun}, {et~al.}}]{G00}
{Ghosh}, K.~K., {Ramsey}, B.~D., {Sadun}, A.~C., {et~al.} 2000, \apj, 537, 638

\bibitem[{{Gu} {et~al.}(2006){Gu}, {Lee}, {Pak}, {et~al.}}]{G06}
{Gu}, M.~F., {Lee}, C.~U., {Pak}, S., {et~al.} 2006, \aap, 450, 39

\bibitem[{{Gupta} {et~al.}(2004){Gupta}, {Banerjee}, {Ashok}, {et~al.}}]{G04}
{Gupta}, A.~C., {Banerjee}, D.~P.~K., {Ashok}, N.~M., {et~al.} 2004, \aap, 422,
  505

\bibitem[{{Gupta} {et~al.}(2008){Gupta}, {Deng}, {Joshi}, {et~al.}}]{G08}
{Gupta}, A.~C., {Deng}, W.~G., {Joshi}, U.~C., {et~al.} 2008, \na, 13, 375

\bibitem[{{Hagen-Thorn} {et~al.}(2002){Hagen-Thorn}, {Larionov}, {Jorstad},
  {et~al.}}]{H02}
{Hagen-Thorn}, V.~A., {Larionov}, V.~M., {Jorstad}, S.~G., {et~al.} 2002, \aj,
  124, 3031

\bibitem[{{Hagen-Thorn} {et~al.}(2004){Hagen-Thorn}, {Larionov}, {Larionova},
  {Kudryavtseva}, {Tikhonov}, {Hagen-Thorn}, {Arkharov}, {di Paola}, \&
  {D'Alessio}}]{H04}
{Hagen-Thorn}, V.~A., {Larionov}, V.~M., {Larionova}, E.~G., {et~al.} 2004,
  AstL, 30, 209

\bibitem[{{Horne} \& {Baliunas}(1986)}]{H86}
{Horne}, J.~H. \& {Baliunas}, S.~L. 1986, \apj, 302, 757

\bibitem[{{Hu} {et~al.}(2011){Hu}, {Wu}, {Guo}, {Zhou}, {Zhang}, \&
  {Zheng}}]{H11}
{Hu}, S.~M., {Wu}, J., {Guo}, H.~Y., {et~al.} 2011, \apss, 333, 213

\bibitem[{{Hyv{\"o}nen} {et~al.}(2007){Hyv{\"o}nen}, {Kotilainen}, {Falomo},
  {et~al.}}]{H07}
{Hyv{\"o}nen}, T., {Kotilainen}, J.~K., {Falomo}, R., {et~al.} 2007, \aap, 476,
  723

\bibitem[{{Jurkevich}(1971)}]{J71}
{Jurkevich}, I. 1971, \apss, 13, 154

\bibitem[{{Katajainen} {et~al.}(2000){Katajainen}, {Takalo},
  {Sillanp{\"a}{\"a}}, {Nilsson}, {Pursimo}, {Hanski}, {Hein{\"a}m{\"a}ki},
  {Kotoneva}, {Lainela}, {Nurmi}, {Pietil{\"a}}, {Rekola}, {Riehokainen},
  {Teerikorpi}, {Valtaoja}, \& {L{\"a}hteenm{\"a}ki}}]{K00}
{Katajainen}, S., {Takalo}, L.~O., {Sillanp{\"a}{\"a}}, A., {et~al.} 2000,
  \aaps, 143, 357

\bibitem[{{Kelly} {et~al.}(2003){Kelly}, {Hughes}, {Aller}, {et~al.}}]{K03}
{Kelly}, B.~C., {Hughes}, P.~A., {Aller}, H.~D., {et~al.} 2003, \apj, 591, 695

\bibitem[{{Maesano} {et~al.}(1997){Maesano}, {Montagni}, {Massaro},
  {et~al.}}]{M97}
{Maesano}, M., {Montagni}, F., {Massaro}, E., {et~al.} 1997, \aaps, 122, 267

\bibitem[{{Marchenko} {et~al.}(1996){Marchenko}, {Hagen-Thorn}, {Yakovleva}, \&
  {Mikolaichuk}}]{M96}
{Marchenko}, S.~G., {Hagen-Thorn}, V.~A., {Yakovleva}, V.~A., \& {Mikolaichuk},
  O.~V. 1996, in ASP Conference Series, Vol. 110, Blazar Continuum Variability,
  ed. H.~R. {Miller}, J.~R. {Webb}, \& J.~C. {Noble}, 105

\bibitem[{{Mead} {et~al.}(1990){Mead}, {Ballard}, {Brand}, {Hough}, {Brindle},
  \& {Bailey}}]{M90}
{Mead}, A.~R.~G., {Ballard}, K.~R., {Brand}, P.~W.~J.~L., {et~al.} 1990, \aaps,
  83, 183

\bibitem[{{Miller} \& {Hawley}(1977)}]{M77}
{Miller}, J.~S. \& {Hawley}, S.~A. 1977, \apjl, 212, L47

\bibitem[{{Paltani} {et~al.}(1997){Paltani}, {Courvoisier}, {Blecha},
  {et~al.}}]{P97}
{Paltani}, S., {Courvoisier}, T.~J.~L., {Blecha}, A., {et~al.} 1997, \aap, 327,
  539

\bibitem[{{Papadakis} {et~al.}(2003){Papadakis}, {Boumis}, {Samaritakis},
  {et~al.}}]{P03}
{Papadakis}, I.~E., {Boumis}, P., {Samaritakis}, V., {et~al.} 2003, \aap, 397,
  565

\bibitem[{{Racine}(1970)}]{R70}
{Racine}, R. 1970, \apjl, 159, L99

\bibitem[{{Raiteri} {et~al.}(2001){Raiteri}, {Villata}, {Aller}, {Aller},
  {Heidt}, {Kurtanidze}, {Lanteri}, {Maesano}, {Massaro}, {Montagni}, {Nesci},
  {Nilsson}, {Nikolashvili}, {Nurmi}, {Ostorero}, {Pursimo}, {Rekola},
  {Sillanp{\"a}{\"a}}, {Takalo}, {Ter{\"a}sranta}, {Tosti}, {Balonek}, {Feldt},
  {Heines}, {Heisler}, {Hu}, {Kidger}, {Mattox}, {McGrath}, {Pati}, {Robb},
  {Sadun}, {Shastri}, {Wagner}, {Wei}, \& {Wu}}]{R01}
{Raiteri}, C.~M., {Villata}, M., {Aller}, H.~D., {et~al.} 2001, \aap, 377, 396

\bibitem[{{Raiteri} {et~al.}(2010){Raiteri}, {Villata}, {Bruschini}, {Capetti},
  {Kurtanidze}, {Larionov}, {Romano}, {Vercellone}, {Agudo}, {Aller}, {Aller},
  {Arkharov}, {Bach}, {Berdyugin}, {Blinov}, {B{\"o}ttcher}, {Buemi},
  {Calcidese}, {Carosati}, {Casas}, {Chen}, {Coloma}, {Diltz}, {di Paola},
  {Dolci}, {Efimova}, {Forn{\'e}}, {G{\'o}mez}, {Gurwell}, {Hakola}, {Hovatta},
  {Hsiao}, {Jordan}, {Jorstad}, {Koptelova}, {Kurtanidze},
  {L{\"a}hteenm{\"a}ki}, {Larionova}, {Leto}, {Lindfors}, {Ligustri},
  {Marscher}, {Morozova}, {Nikolashvili}, {Nilsson}, {Ros}, {Roustazadeh},
  {Sadun}, {Sillanp{\"a}{\"a}}, {Sainio}, {Takalo}, {Tornikoski}, {Trigilio},
  {Troitsky}, \& {Umana}}]{R10}
{Raiteri}, C.~M., {Villata}, M., {Bruschini}, L., {et~al.} 2010, \aap, 524, A43

\bibitem[{{Raiteri} {et~al.}(2009){Raiteri}, {Villata}, {Capetti}, {Aller},
  {Bach}, {Calcidese}, {Gurwell}, {Larionov}, {Ohlert}, {Nilsson},
  {Strigachev}, {Agudo}, {Aller}, {Bachev}, {Ben{\'{\i}}tez}, {Berdyugin},
  {B{\"o}ttcher}, {Buemi}, {Buttiglione}, {Carosati}, {Charlot}, {Chen},
  {Dultzin}, {Forn{\'e}}, {Fuhrmann}, {G{\'o}mez}, {Gupta}, {Heidt}, {Hiriart},
  {Hsiao}, {Jel{\'{\i}}nek}, {Jorstad}, {Kimeridze}, {Konstantinova},
  {Kopatskaya}, {Kostov}, {Kurtanidze}, {L{\"a}hteenm{\"a}ki}, {Lanteri},
  {Larionova}, {Leto}, {Latev}, {Le Campion}, {Lee}, {Ligustri}, {Lindfors},
  {Marscher}, {Mihov}, {Nikolashvili}, {Nikolov}, {Ovcharov}, {Principe},
  {Pursimo}, {Ragozzine}, {Robb}, {Ros}, {Sadun}, {Sagar}, {Semkov}, {Sigua},
  {Smart}, {Sorcia}, {Takalo}, {Tornikoski}, {Trigilio}, {Uckert}, {Umana},
  {Valcheva}, \& {Volvach}}]{R09}
{Raiteri}, C.~M., {Villata}, M., {Capetti}, A., {et~al.} 2009, \aap, 507, 769

\bibitem[{{Raiteri} {et~al.}(2013){Raiteri}, {Villata}, {D'Ammando},
  {Larionov}, {Gurwell}, {Mirzaqulov}, {Smith}, \& {Carnerero}}]{R13}
{Raiteri}, C.~M., {Villata}, M., {D'Ammando}, F., {et~al.} 2013, in European
  Physical Journal Web of Conferences, Vol.~61, 4014

\bibitem[{{Raiteri} {et~al.}(2003){Raiteri}, {Villata}, {Tosti}, {Nesci},
  {Massaro}, {Aller}, {Aller}, {Ter{\"a}sranta}, {Kurtanidze}, {Nikolashvili},
  {Ibrahimov}, {Papadakis}, {Krichbaum}, {Kraus}, {Witzel}, {Ungerechts},
  {Lisenfeld}, {Bach}, {Cim{\`o}}, {Ciprini}, {Fuhrmann}, {Kimeridze},
  {Lanteri}, {Maesano}, {Montagni}, {Nucciarelli}, \& {Ostorero}}]{R03}
{Raiteri}, C.~M., {Villata}, M., {Tosti}, G., {et~al.} 2003, \aap, 402, 151

\bibitem[{{Rieger}(2004)}]{R04}
{Rieger}, F.~M. 2004, \apjl, 615, L5

\bibitem[{{Simonetti} {et~al.}(1985){Simonetti}, {Cordes}, \& {Heeschen}}]{S85}
{Simonetti}, J.~H., {Cordes}, J.~M., \& {Heeschen}, D.~S. 1985, \apj, 296, 46

\bibitem[{{Smith} \& {Nair}(1995)}]{S95}
{Smith}, A.~G. \& {Nair}, A.~D. 1995, \pasp, 107, 863

\bibitem[{{Smith} {et~al.}(2009){Smith}, {Montiel}, {Rightley}, {Turner},
  {Schmidt}, \& {Jannuzi}}]{S09}
{Smith}, P.~S., {Montiel}, E., {Rightley}, S., {et~al.} 2009, arXiv: 0912.3621,
  2009 Fermi Symposium, eConf Proceedings C091122

\bibitem[{{Sobrito} {et~al.}(1999){Sobrito}, {Villata}, {Raiteri}, {de
  Francesco}, {Lanteri}, \& {Cavallone}}]{S99}
{Sobrito}, G., {Villata}, M., {Raiteri}, C.~M., {et~al.} 1999, BLAZAR Data, 1,
  5

\bibitem[{{Tosti} {et~al.}(1999){Tosti}, {Luciani}, {Fiorucci}, {Massaro},
  {Nesci}, {Maesano}, {Montagni}, {Villata}, {Raiteri}, {de Francesco},
  {Sobrito}, {Takalo}, {Sillanp{\"a}{\"a}}, {Katajainen}, {Pursimo},
  {Kurtanidze}, \& {Nikolashvili}}]{T99}
{Tosti}, G., {Luciani}, M., {Fiorucci}, M., {et~al.} 1999, \memsai, 70, 237

\bibitem[{{Vagnetti} {et~al.}(2003){Vagnetti}, {Trevese}, \& {Nesci}}]{V03}
{Vagnetti}, F., {Trevese}, D., \& {Nesci}, R. 2003, \apj, 590, 123

\bibitem[{{Villata} \& {Raiteri}(1999)}]{V99}
{Villata}, M. \& {Raiteri}, C.~M. 1999, \aap, 347, 30

\bibitem[{{Villata} {et~al.}(2004{\natexlab{a}}){Villata}, {Raiteri}, {Aller},
  {Aller}, {Ter{\"a}sranta}, {Koivula}, {Wiren}, {Kurtanidze}, {Nikolashvili},
  {Ibrahimov}, {Papadakis}, {Tosti}, {Hroch}, {Takalo}, {Sillanp{\"a}{\"a}},
  {Hagen-Thorn}, {Larionov}, {Schwartz}, {Basler}, {Brown}, \&
  {Balonek}}]{V04b}
{Villata}, M., {Raiteri}, C.~M., {Aller}, H.~D., {et~al.} 2004{\natexlab{a}},
  \aap

\bibitem[{{Villata} {et~al.}(2004{\natexlab{b}}){Villata}, {Raiteri},
  {Kurtanidze}, {Nikolashvili}, {Ibrahimov}, {Papadakis}, {Tosti}, {Hroch},
  {Takalo}, {Sillanp{\"a}{\"a}}, {Hagen-Thorn}, {Larionov}, {Schwartz},
  {Basler}, {Brown}, {Balonek}, {Ben{\'{\i}}tez}, {Ram{\'{\i}}rez}, {Sadun},
  {Boltwood}, {Carini}, {Barnaby}, {Coloma}, {Ros}, {Dai}, {Xie}, {Mattox},
  {Rodriguez}, {Asfandiyarov}, {Atkerson}, {Beem}, {Bloom}, {Chanturiya},
  {Ciprini}, {Crapanzano}, {de Diego}, {Efimova}, {Gardiol}, {Guerra},
  {Kahharov}, {Kapanadze}, {Karttunen}, {Kato}, {Kimeridze}, {Kudryavtseva},
  {Lainela}, {Lanteri}, {Larionova}, {Maesano}, {Marchili}, {Massone},
  {Monroe}, {Montagni}, {Nesci}, {Nilsson}, {Noble}, {Nucciarelli}, {Ostorero},
  {Papamastorakis}, {Pasanen}, {Peters}, {Pursimo}, {Reig}, {Ryle}, {Sclavi},
  {Sigua}, {Uemura}, \& {Wills}}]{V04a}
{Villata}, M., {Raiteri}, C.~M., {Kurtanidze}, O.~M., {et~al.}
  2004{\natexlab{b}}, \aap

\bibitem[{{Villata} {et~al.}(2002){Villata}, {Raiteri}, {Kurtanidze},
  {Nikolashvili}, {Ibrahimov}, {Papadakis}, {Tsinganos}, {Sadakane}, {Okada},
  {Takalo}, {Sillanp{\"a}{\"a}}, {Tosti}, {Ciprini}, {Frasca}, {Marilli},
  {Robb}, {Noble}, {Jorstad}, {Hagen-Thorn}, {Larionov}, {Nesci}, {Maesano},
  {Schwartz}, {Basler}, {Gorham}, {Iwamatsu}, {Kato}, {Pullen},
  {Ben{\'{\i}}tez}, {de Diego}, {Moilanen}, {Oksanen}, {Rodriguez}, {Sadun},
  {Kelly}, {Carini}, {Miller}, {Catalano}, {Dultzin-Hacyan}, {Fan}, {Ishioka},
  {Karttunen}, {Kein{\"a}nen}, {Kudryavtseva}, {Lainela}, {Lanteri},
  {Larionova}, {Matsumoto}, {Mattox}, {Montagni}, {Nucciarelli}, {Ostorero},
  {Papamastorakis}, {Pasanen}, {Sobrito}, \& {Uemura}}]{V02}
{Villata}, M., {Raiteri}, C.~M., {Kurtanidze}, O.~M., {et~al.} 2002, \aap, 390,
  407

\bibitem[{{Villata} {et~al.}(2009){Villata}, {Raiteri}, {Larionov},
  {Nikolashvili}, {Aller}, {Bach}, {Carosati}, {Hroch}, {Ibrahimov}, {Jorstad},
  {Kovalev}, {L{\"a}hteenm{\"a}ki}, {Nilsson}, {Ter{\"a}sranta}, {Tosti},
  {Aller}, {Arkharov}, {Berdyugin}, {Boltwood}, {Buemi}, {Casas}, {Charlot},
  {Coloma}, {di Paola}, {di Rico}, {Kimeridze}, {Konstantinova}, {Kopatskaya},
  {Kovalev}, {Kurtanidze}, {Lanteri}, {Larionova}, {Larionova}, {Le Campion},
  {Leto}, {Lindfors}, {Marscher}, {Marshall}, {McFarland}, {McHardy}, {Miller},
  {Nucciarelli}, {Osterman}, {Pasanen}, {Pursimo}, {Ros}, {Sadun}, {Sigua},
  {Sixtova}, {Takalo}, {Tornikoski}, {Trigilio}, {Umana}, {Xie}, {Zhang}, \&
  {Zhou}}]{V09}
{Villata}, M., {Raiteri}, C.~M., {Larionov}, V.~M., {et~al.} 2009, \aap, 501,
  455

\bibitem[{{Villata} {et~al.}(1998){Villata}, {Raiteri}, {Sillanpaa}, \&
  {Takalo}}]{V98}
{Villata}, M., {Raiteri}, C.~M., {Sillanpaa}, A., \& {Takalo}, L.~O. 1998,
  \mnras, 293, L13

\bibitem[{Wang(2014)}]{W14}
Wang, H. 2014, SCPMA, 57, 375

\bibitem[{{Webb} {et~al.}(1998){Webb}, {Freedman}, {Howard}, {Ma}, {Belfort},
  {Rave}, {Rumstay}, {Nicol}, {Krick}, {Oswalt}, {Marshall}, \&
  {Robishaw}}]{W98}
{Webb}, J.~R., {Freedman}, I., {Howard}, E., {et~al.} 1998, \aj, 115, 2244

\bibitem[{{Wu} {et~al.}(2007){Wu}, {Zhou}, {Ma}, {Wu}, {Jiang}, \&
  {Chen}}]{W07}
{Wu}, J., {Zhou}, X., {Ma}, J., {et~al.} 2007, \aj, 133, 1599

\bibitem[{{Zhang} {et~al.}(2004){Zhang}, {Zhang}, {Zhao}, {Xie}, {Wu}, \&
  {Zheng}}]{Z04}
{Zhang}, X., {Zhang}, L., {Zhao}, G., {et~al.} 2004, \aj, 128, 1929

\bibitem[{{Zhang} {et~al.}(2013){Zhang}, {Bian}, {Li}, \& {Shang}}]{Z13}
{Zhang}, Y.~H., {Bian}, F.~Y., {Li}, J.~Z., \& {Shang}, R.~C. 2013, \mnras,
  432, 1189

\end{thebibliography}
\end{document}